\documentclass[10pt]{iopart}

\usepackage{graphicx}
\usepackage{amssymb}
\usepackage{iopams}
\usepackage{amsthm}
\usepackage{latexsym}
\usepackage{fontenc}
\usepackage{geometry}
\usepackage{setspace}

\newcommand{\I}{{\rm i}}

\newcommand{\be}{\begin{equation}}
\newcommand{\ee}{\end{equation}}
\newcommand{\bea}{\begin{eqnarray}}
\newcommand{\eea}{\end{eqnarray}}

\def\J#1#2#3#4{#1 {\it #2} {\bf #3} #4}

\def\PR{\it Phys. Rev.}

\def\JMP{\it J. Math. Phys.}
\def\CQG{\it Class. Quantum Grav.}

\def\PLA{\it Phys. Lett. A}

\begin{document}

\title [About Poynting vector in axially symmetric stationary 
spacetimes]{\textbf {About Poynting vector in axially symmetric stationary 
spacetimes }}

\author {
                Leonardo A. Pach\'{o}n$^{1}$
                \footnote[1]{e-mail: leaupaco@laft.org}
                and
                Jorge A. Rueda$^{2,3}$
                \footnote[2]{e-mail: jrueda@ula.ve}
                }
\address{
         $^{1}$ Departamento de F\'isica, Universidad Nacional de Colombia, Santa F\'e de Bogot\'a D. C., 
                Colombia.\\
         $^{2}$ Escuela de F\'{i}sica, Universidad Industrial de Santander, Bucaramanga, Colombia.\\
         $^{3}$ Centro de F\'isica Te\'orica, Universidad de Los Andes, M\'erida, Venezuela.
         }
\begin{abstract}
We dispose of some objections raised  by Manko \emph{et 
al.} \cite{Manko} on a recently published paper on the role of Poynting vector in the ocurrence of vorticity in electrovaccum spacetimes \cite{Herrera}.
\end{abstract}

In order to confirm the Bonnor's hyphothesis about the role of the electromagnetic 
energy flux as the responsible for  the vorticity in the surrounding of non--rotating sources of 
some gravitational fields \cite{Bon}, Herrera \emph{et al.} \cite{Herrera} recently presented 
a general study on the vorticity  in axially 
symmetric stationary electrovacuum spacetimes, considering the influence of, both, the 
mass rotations and electromagnetic fields.

As was pointed out in \cite{Manko}, the authors of \cite{Herrera} define the electromagnetic 
Ernst potential as $\Phi(\rho,z)=\phi + i \psi$ and the electromagnetic 4-potential as 
$A_{\mu}=(\phi,0,A,0)$ forgetting to change the sign of the  time component of $A_{\mu}$. 
This misprint is translated to the definition of  the Poynting vector, 
\footnote{The missing  4 in the denominator of $S{^\phi}$ in \cite{Herrera} is a typographic misprint without any relevance.}
which should read 
\begin{equation}\label{poyntingVector}
S{^\phi}= \frac{f^{3/2}}{4\pi \rho^2\,e^{2\gamma}}\left(\omega \nabla
\phi \cdot \nabla \phi - \nabla\phi\cdot \nabla A\right)\, .
\end{equation}
instead of 
\begin{equation}\label{poyntingVector}
S{^\phi}= \frac{f^{3/2}}{4\pi \rho^2\,e^{2\gamma}}\left(\omega \nabla
\phi \cdot \nabla \phi + \nabla\phi\cdot \nabla A\right)\, .
\end{equation}

The above mentioned  fact, the wrong sign in the definition of the electromagnetic four potential, 
is the leiv motiv of the paper by Manko \emph{et al.} \cite{Manko}.

All this having been said, the following comments are in order:
\begin{itemize}

\item  All the physical conclusions in \cite{Herrera} are  correct, and remain the same independently on the misprint mentioned before.

\item The comment in  \cite{Manko} ``....we observe that in \cite{Herrera} the electrostatic limit 
of Manko's solution was accomplished ...... for getting the factor $\I qb...$'' , is irrelevant, and might even be misleading for a  non attentive reader.
Indeed,  equation (39) in \cite{Herrera} was obtained as follows: the $\Omega$ potential was quoted from  \cite{Ruiz} and then the limit 
for nor arbitrary cuadripolar deformation ($k\rightarrow 0$) and for rigid rotations ($b \rightarrow 0$) was taken. 
\end{itemize}

\ack

The authors wish to thank  Professor Luis Herrera for useful comments and for checking the redaction of this note.

\end{document}